# New self-dual codes of length 72

Alexandre Zhdanov
Voronezh, Russia
a-zhdan@vmail.ru

*Abstract—* In this paper we obtain at least 61 new singly even (Type I) binary [72,36,12] self-dual codes as a quasi-cyclic codes with m=2 (tailbitting convolutional codes) and at least 13 new doubly even (Type II) binary [72,36,12] self-dual codes by replacing the first row in each circulant in a double circulant code by "all ones" and "all zeros" vectors respectively.

## I. Introduction

A linear binary code $C(n,k,d)$ is a subspace of $F_2^n$ of dimension $k$. The $F_2$ is a field of two elements: 1,0, where the summation is a logical XOR and multiplication is a logical AND. The $F_2^n$ is n-tuple of $F_2$. The codeword weight $d$ is a minimal number of non-zero component in any codeword of code $C$. The quasi-cyclic code is a code for which every cyclic shift of a codeword by $m$ symbols yields another valid codeword, where $m > 1$. The quasi-cyclic code of $R = 1/m$ consists of $m$ circulants. A circulant is a square matrix where the next row is obtained by one element cyclically shifting to the right the previous row. The cyclically shifting to the left will result an inverse circulant. The tailbitting convolutional code of $R = 1/2$ is a quasi-cyclic code with $m = 2$, where the columns of the circulants are mixed to form a compact mixed polynomial string. The mixed polynomial string is a non-zero part of the generator matrix row. Self dual codes are a powerful class of codes. Self-dual code $C$ is a code with coding rate $R = 1/2$, where the inner product of any two rows in a generator matrix $G$ gives 0. In other words: $C = C^\perp$, where $C^\perp$ is a dual code. All codeword's of binary self-dual code has even weight. If all codewords weights $\equiv 0 \pmod{4}$ the code is called doubly even, if all codewords weights $\equiv 2 \pmod{4}$ the code is called singly even. The code is called extremal if the minimum weight of the codeword meets the following bond: $d \leq 4\lfloor n/24 \rfloor + 6$ if $n \equiv 22 \pmod{24}$ and $d \leq 4 \lfloor n/24 \rfloor + 4$ otherwise. We refer the reader to [1] for details.

Let us consider the convolutional codes and its taps are described by the polynomials of constraint length $K$ (Type $A_0$ [2]). In this case the generator matrix is obtained for example by cyclically shift of the mixed polynomial string $(p_0, q_0, p_1, q_1, \cdots, p_{K-1}, q_{K-1}, 0, 0, \cdots 0)$ with step 2 or use another form of generator matrix $G = [P | Q]$, where $P$ and $Q$ are circulants $k \times k$ with top row $(p_0, p_1, \cdots, p_{K-1}, 0, 0, \cdots 0)$ and $(q_0, q_1, \cdots, q_{K-1}, 0, 0, \cdots 0)$ respectively. Note that the generator matrix produced by two circulants: the forward and the inverse with the same first row is a generator matrix of a self-dual code. The standard inner product between the first and second row in the first circulant will be: $a = x_0 x_k + x_1 x_0 + x_2 x_1 + \cdots + x_k x_{k-1}$ and in the inverse circulant the result will be $a_{inv} = x_0 x_1 + x_1 x_2 + \cdots + x_{k-1} x_k + x_k x_0$. The $a = a_{inv}$ and the resulting sum will be $0$. This is true for any possible shifts. So, the polynomial pair $P_1 = [p_0, p_1, p_2, \ldots, p_{K-1}]$ and $P_2 = [p_{K-1}, p_{K-2}, \ldots, p_0]$ could be used for convolutional self-dual code generation. Further we will point out only the first polynomial. The second one will be obtained by inversing the first.

When one circulant is an identity matrix $I$ the construction $G = [I | F]$ is called pure double circulant [3]. The connection between quasi-cyclic and pure double circulant is established by theorem 1.3 from [2]. The code C generated by $G = [P|Q]$ can also be generated by $G = [I | F]$, where $F$ is a circulant, iff $gcd(p(x), x^k - 1) = 1$. In such case the connection is $q(x) = p(x) f(x) \mod(x^k - 1)$.

The theorem 1.1 from [2] established that $rank[P|Q] = k - \deg(\gcd(p(x), q(x), x^k - 1))$ in other words to avoid zero-weight codeword must satisfy $\gcd(p(x), q(x), x^k - 1) = 1$.

The possible weight enumerators for singly even self-dual codes of length 72 are given in [5] as

$$W'_{72,1} = 1 + 2\beta y^{12} + (8640 - 64\lambda) y^{14} \\ + (124281 - 24\beta + 384\gamma) y^{16} + \cdots$$

and

$$W'_{72,2} = 1 + 2\beta y^{12} + (7616 - 64\gamma) y^{14} + \\ (134521 - 24\beta + 384\gamma) y^{16} + \cdots$$

The possible weight enumerator for doubly even self-dual codes of length 72 is given in [11] as

$$W''_{72} = 1 + (4398 + \alpha) y^{12} + (197073 - 12\alpha) y^{16} + \\ (18396972 + 66\alpha) y^{20} + \cdots$$

Our database of singly even [72, 36, 12] self-dual codes includes 1088 codes from [4-9].

Our database of doubly even [72, 36, 12] self-dual codes includes 429 codes from [5-12].

In [13] we find a new singly even [72,36,12] code with weight enumerator $W_{72,1}$ and parameters $\beta = 483$, $\gamma = 0$ and develop a new construction (Type $A_3$) for doubly even self-dual convolutional codes of length $n \equiv 0 \pmod 4$. Both based on quasi-cyclic codes with forward and reverse circulants with the same first row. The aim of this paper is to find a new codes through exhaustive search a good polynomial of any possible constraint length where the generators polynomials will be in the following form: $P_1 = [p_0, p_1, p_2, \ldots, p_{K-1}]$ and $P_2 = [p_{K-1}, p_{K-2}, \ldots, p_0]$.

## II. CODE CONSTRUCTION

Let us note, that $x^{36} - 1$ has two divisors: $x^2 - x - 1$ and $x - 1$. For singly even codes (each polynomial has odd weight) we are able to satisfy theorem 1.3 from [2] and construct code which could be represented as pure double circulant i. e. $gcd(P_1(x), x^{36} - 1) = 1$ and $gcd(P_2(x), x^{36} - 1) = 1$. It is not possible for doubly even codes where each polynomial has even weight. The minimal possible greatest common divisor for even weight polynomials will be $gcd(P_1(x), x^{36} - 1) = x - 1$ and $gcd(P_2(x), x^{36} - 1) = x - 1$ because polynomial with even number of ones has divisor $x - 1$. In such case we will use Type $A_3$ construction from [13] which includes the following. In the forward circulant we replace the first row with "all zeros" vector and in the reverse circulant we replace the first row with "all ones" vector. Then we verify this code by the full weight enumerator analysis. The obtained doubly even self-dual codes are not pure double circulants. All doubly even pure double circulants of length 72 are listed in [10]. The replaced row is a linear combination of the other rows so the obtained code is a quasi-cyclic code with $m = 2$.

We provide the exhaustive search for polynomials that satisfy the given conditions.

## III. MAIN RESULT

### A. New singly even self-dual codes of length 72

We have obtained 307 singly even self-dual codes with minimal weight codeword $d = 12$. The codes have the 83 unique combinations of parameters $\gamma$ and $\beta$. The 215 codes are the codes with previously unknown values of $\gamma, \beta$, the 61 codes of them have unique parameters. The codes with previously unknown parameters (unique $\gamma$ and $\beta$) and minimal possible constrained length of the generator polynomials are listed in the Table 1. All codes have weight enumerator $W'_{72,1}$.

TABLE I. SINGLY EVEN [72,36,12] CODES WITH NEW $\gamma$ AND $\beta$

| N | $\gamma$ | $\beta$ | P | K | Number of ones |
|---|---|---|---|---|---|
| 1 | 0 | 483 | 1111001011 | 10 | 7 |
| 2 | 0 | 330 | 11110110001 | 11 | 7 |
| 3 | 0 | 324 | 11110010011 | 11 | 7 |
| 4 | 0 | 294 | 111101001001 | 12 | 7 |
| 5 | 36 | 765 | 111100110001 | 12 | 7 |
| 6 | 0 | 285 | 111010001101 | 12 | 7 |
| 7 | 0 | 162 | 110101000111 | 12 | 7 |
| 8 | 0 | 171 | 110011100101 | 12 | 7 |
| 9 | 0 | 360 | 110010011011 | 12 | 7 |
| 10 | 0 | 423 | 110001110011 | 12 | 7 |
| 11 | 36 | 576 | 111110110101 | 12 | 9 |
| 12 | 36 | 642 | 111011110011 | 12 | 9 |
| 13 | 36 | 528 | 1110000110101 | 13 | 7 |
| 14 | 0 | 288 | 1110000100111 | 13 | 7 |
| 15 | 0 | 183 | 1101101010001 | 13 | 7 |
| 16 | 36 | 630 | 1101011000101 | 13 | 7 |
| 17 | 0 | 312 | 1101010011001 | 13 | 7 |
| 18 | 0 | 396 | 1101001000111 | 13 | 7 |
| 19 | 0 | 318 | 1101000010111 | 13 | 7 |
| 20 | 0 | 177 | 1100100010111 | 13 | 7 |
| 21 | 0 | 306 | 1111100011101 | 13 | 9 |
| 22 | 0 | 405 | 1100010011111 | 13 | 9 |
| 23 | 0 | 180 | 1101110011101 | 13 | 9 |
| 24 | 0 | 393 | 1101100111101 | 13 | 9 |
| 25 | 0 | 147 | 1101100101111 | 13 | 9 |
| 26 | 36 | 492 | 1111101110111 | 13 | 11 |
| 27 | 0 | 141 | 1111011011111 | 13 | 11 |
| 28 | 36 | 519 | 1111010111111 | 13 | 11 |
| 29 | 0 | 432 | 11010000110101 | 14 | 7 |
| 30 | 0 | 204 | 11000101110001 | 14 | 7 |
| 31 | 0 | 348 | 11000011011001 | 14 | 7 |
| 32 | 0 | 114 | 11000010011011 | 14 | 7 |
| 33 | 36 | 513 | 10111001100001 | 14 | 7 |
| 34 | 0 | 87 | 11111001101001 | 14 | 9 |
| 35 | 36 | 327 | 11101110100101 | 14 | 9 |
| 36 | 36 | 369 | 11101101110001 | 14 | 9 |
| 37 | 0 | 153 | 11100101001111 | 14 | 9 |

| N | $\gamma$ | $\beta$ | P | K | Number of ones |
|---|---|---|---|---|---|
| 38 | 72 | 825 | 111000010111111 | 14 | 9 |
| 39 | 36 | 357 | 110110010101111 | 14 | 9 |
| 40 | 0 | 276 | 111110111010111 | 14 | 11 |
| 41 | 0 | 735 | 111100100010001 | 15 | 7 |
| 42 | 36 | 201 | 111010100010001 | 15 | 7 |
| 43 | 0 | 96 | 111000010110001 | 15 | 7 |
| 44 | 36 | 438 | 110010101100001 | 15 | 7 |
| 45 | 0 | 189 | 110001110000101 | 15 | 7 |
| 46 | 36 | 465 | 101101101000001 | 15 | 7 |
| 47 | 0 | 504 | 111111001010001 | 15 | 9 |
| 48 | 36 | 498 | 111100110101001 | 15 | 9 |
| 49 | 0 | 402 | 111100011001101 | 15 | 9 |
| 50 | 36 | 363 | 111100001111001 | 15 | 9 |
| 51 | 0 | 387 | 111011100101001 | 15 | 9 |
| 52 | 0 | 351 | 111010101011001 | 15 | 9 |
| 53 | 0 | 159 | 111001011110001 | 15 | 9 |
| 54 | 36 | 537 | 111001001011101 | 15 | 9 |
| 55 | 0 | 252 | 110111100010101 | 15 | 9 |
| 56 | 0 | 333 | 110101100001111 | 15 | 9 |
| 57 | 0 | 417 | 111111010011101 | 15 | 11 |
| 58 | 0 | 240 | 1110001000011001 | 16 | 7 |
| 59 | 0 | 195 | 1101010010001001 | 16 | 7 |
| 60 | 36 | 783 | 1100010100110001 | 16 | 7 |
| 61 | 36 | 531 | 1011001100010001 | 16 | 7 |

Fig. 1. The generator matrix G of self-dual tailbitting convolutional code for n=72, k=36, P=(1111001011)₂ with $\beta = 483, \gamma = 0$.

The mixed polynomial string (non-zero part of the first row) is q=[1,1,1,1,1,0,1,1,0,0,0,0,1,1,0,1,1,1,1,1]. The odd positions are occupied by $P=[1,1,1,1,0,0,1,0,1,1]$ and the even positions are occupied by its inverse $[1,1,0,1,0,0,1,1,1,1]$. The obtained codes are the first known codes with $\gamma = 36$ and $\beta = 765, 576, 642, 528, 630, 492, 519, 513, 327, 369, 357, 201, 438, 465, 498, 363, 537, 783, 531$; with $\gamma = 0$ and $\beta = 483, 330, 324, 294, 285, 162, 171, 360, 423, 288, 183, 312, 396, 318, 177, 306, 405, 180, 393, 147, 141, 432, 204, 348, 114, 87, 153, 276, 735, 96, 189, 504, 402, 387, 351, 159, 252, 333, 417, 240, 195$.

*B. New doubly even self-dual codes of length 72*

We have obtained 818 doubly even self-dual codes with minimal weight codeword $d=12$. The codes have the 79 unique $\alpha$ values. The 110 codes have the 13 new unique values of $\alpha$. The codes with unique $\alpha$ and minimal possible constrained length of the generator polynomials are listed in the Table 2. All codes have weight enumerator $W''_{72}$.

Fig. 2. The generator matrix G of self-dual code for n=72, k=36, P=(11100110111)₂ with $\alpha = -2424$.

TABLE II. DOUBLY EVEN [72,36,12] CODES WITH NEW $\alpha$

| N | $\alpha$ | P | K | Number of ones |
|---|---|---|---|---|
| 1 | -2424 | 11100110111 | 11 | 8 |
| 2 | -2784 | 11110001001 | 11 | 6 |
| 3 | -2712 | 10111010001 | 11 | 6 |
| 4 | -2580 | 111010111001 | 12 | 8 |
| 5 | -2832 | 1111100111011 | 13 | 10 |
| 6 | -2532 | 1111011101101 | 13 | 10 |
| 7 | -2964 | 1110010110101 | 13 | 8 |
| 8 | -3060 | 11100011111101 | 14 | 10 |
| 9 | -2988 | 110111111000001 | 14 | 8 |
| 10 | -2736 | 111000011110001 | 15 | 8 |
| 11 | -3048 | 111111110100001 | 15 | 10 |
| 12 | -2844 | 111110100011101 | 15 | 10 |
| 13 | -3132 | 110111110001101 | 15 | 10 |

The obtained codes are the first known codes with $\alpha$ = -2424, -2784, -2712, -2580, -2832, -2532, -2964, -3060, -2988, -2736, -3048, -2844, -3132.

## IV. CONCLUSION

We have obtained at least 61 singly even self-dual [72,36,12] codes with new unique $\gamma$ and $\beta$ parameters. All codes have weight enumerator $W'_{72,1}$. The obtained codes are the first known codes with $\gamma = 36$ and $\beta$ = 765, 576, 642, 528, 630, 492, 519, 513, 327, 369, 357, 201, 438, 465, 498, 363, 537, 783, 531; with $\gamma = 0$ and $\beta$ = 483, 330, 324, 294, 285, 162, 171, 360, 423, 288, 183, 312, 396, 318, 177, 306, 405, 180, 393, 147, 141, 432, 204, 348, 114, 87, 153, 276, 735, 96, 189, 504, 402, 387, 351, 159, 252, 333, 417, 240, 195. It is notable that a number of different codes may have the same $\gamma$ and $\beta$ parameters. We list the obtained singly even [72,36,12] self-dual codes with unique $\gamma$ and $\beta$ and minimal constraint length of the generator polynomials in Appendix A Table IV. The string format is given below the table. We designate as known only the (36 441 14 7 2A4B) code. It is a same as the code D18 from [4].

We have obtained at least 13 doubly even self-dual [72,36,12] codes with new unique values of $\alpha$. All codes have weight enumerator $W''_{72}$. The obtained codes are the first known codes with $\alpha$ = -2424, -2784, -2712, -2580, -2832, -2532, -2964, -3060, -2988, -2736, -3048, -2844, -3132.

We list the obtained doubly even [72,36,12] self-dual codes with unique $\alpha$ parameter and minimal possible constraint length of the generator polynomials in Appendix A Table III. It is notable that the obtained codes are not pure or bordered double circulants but may have the same weight enumerator.

## REFERENCES

[1] Gabriele Nebe, Eric M. Rains, and Neil J. A. Sloane. 2006. Self-Dual Codes and Invariant Theory (Algorithms and Computation in Mathematics). Springer-Verlag New York, Inc., Secaucus, NJ, USA.

[2] Solomon G., van Tilborg H.C.A., "A connection between block and convolutional codes," SIAM J. Applied Math., V37, N2, pp. 358-369, 10/79.

[3] W.C. Huffman and V. Pless, Fundamentals of Error Correcting Codes, Cambridge University Press, 2003.

[4] J.H. Conway and N.J.A. Sloane, "A new upper bound on the minimum distance of self-dual codes", IEEE Trans. Inform. Theory, 36, (1990), p. 1319-1333

[5] A. Kaya, B. Yildiz and I. Siap, "New extremal binary self-dual codes of length 68 from quadratic residue codes over F2 + uF2 + u2F2", arXiv:1308.0580, 2013.

[6] S.T. Dougherty, J-L. Kim and P. Sole, "Double circulant codes from two class association schemes", Advances in Mathematics of Communications, vol. 1, no.1, pp. 45–64, 2007.

[7] N. Yankov, M.H. Lee, M. Gürel and M. Ivanova, Self-dual codes with an automorphism of order 11, IEEE Trans. Inform. Theory 61 (2015), 1188–1193.

[8] Gürel, M., Yankov, N. (2016). Self-dual codes with an automorphism of order 17. Mathematical Communications, 21(1), 97-107. Retrieved from http://hrcak.srce.hr/157710

[9] Nesibe Tufekci, Bahattin Yildiz, "On codes over $R_{k,m}$ and constructions for new binary self-dual", arXiv: 1406.1281, 2014

[10] T. A. Gulliver, M. Harada, On doubly circulant doubly even self-dual [72, 36, 12] codes and their neighbors, Australas. J. Combin. 40(2008), 137–144.

[11] S.T. Dougherty, T.A. Gulliver, M. Harada, "Extremal binary self dual codes", IEEE Trans.Infrom. Theory, Vol. 43, No.6, pp. 2036–2047, 1997.

[12] Iliya Bouyukliev and Veerle Fack and Joost Winne (2005), Hadamard matrices of order 36, in 2005 European Conference on Combinatorics, Graph Theory and Applications (EuroComb '05), Stefan Felsner (ed.), Discrete Mathematics and Theoretical Computer Science Proceedings AE, pp. 93-98

[13] A. Zhdanov, "Convolutional encoding of 60,64,68,72-bit self-dual codes", arXiv:1702.05153, 2017

## APPENDIX A

TABLE III. LIST OF OBTAINED DOUBLY EVEN [72,36,12] SELF-DUAL CODES WITH UNIQUE ALPHAS AND MINIMAL CONSTRAINT LENGTH*

| | | | |
|---|---|---|---|
| -2748 9 6 12F | -2580 12 8 9D7 | -3756 13 8 147B | -3084 14 8 3C2B |
| -2820 9 6 1CB | -3624 12 8 AE7 | -3180 13 8 1B2B | -3564 14 8 2F0B |
| -2544 10 8 3D7 | -2976 12 8 E9B | -3192 13 6 109B | -2940 14 8 3C53 |
| -3108 10 6 24F | -3552 12 8 CEB | -3372 13 6 10B3 | -3300 15 8 509F |
| -2856 10 6 2D3 | -3000 12 8 F2B | -3228 13 6 111D | -2652 15 8 5297 |
| -3588 11 8 4DF | -3492 12 6 84F | -3024 13 6 1075 | -2736 15 8 4787 |
| -3336 11 8 65F | -2916 12 6 90F | -3384 14 10 25BF | -3720 15 8 42BB |
| -3120 11 8 5E7 | -3324 12 6 897 | -3288 14 10 29DF | -3600 15 8 7253 |
| -2424 11 8 767 | -2772 12 6 94B | -3432 14 10 336F | -2904 15 8 7D03 |
| -2784 11 6 48F | -3036 12 6 8D3 | -3312 14 10 2EAF | -3048 15 10 42FF |
| -3144 11 6 517 | -3360 12 6 86D | -2676 14 10 39CF | -3072 15 10 43BF |
| -2928 11 6 43B | -3408 13 10 19BF | -3012 14 10 2CF7 | -2844 15 10 5C5F |
| -3276 11 6 46B | -3708 13 10 1ADF | -3060 14 10 2FC7 | -3744 15 10 4DAF |
| -2640 11 6 4CB  | -3456 13 10 1D5F | -3792 14 10 25FB | -2892 15 10 7567 |
| -3468 11 6 4E3  | -2832 13 10 1B9F | -3480 14 10 27DB | -3204 15 10 7E87 |
| -2712 11 6 45D  | -2532 13 10 16EF | -3396 14 8 223F  | -3132 15 10 58FB |
| -2808 12 8 8BF  | -2880 13 10 1AF7 | -3420 14 8 283F  | -2868 16 8 85C7  |
| -3168 12 8 C5F  | -3504 13 8 119F  | -3216 14 8 28E7  | -3696 16 8 D28B  |
| -3444 12 8 8EF  | -3348 13 8 12D7  | -2988 14 8 20FB  | -3828 16 8 91D3  |
| -3156 12 8 C77  | -2964 13 8 15A7  | -3240 14 8 2D2B  | |

*The string format: Alpha, K, constraint length, number of ones, polynomial in hexadecimal form

TABLE IV. LIST OF OBTAINED SINGLY EVEN [72,36,12] SELF-DUAL CODES WITH UNIQUE GAMMA AND BETA AND MINIMAL CONSTRAINT LENGTH*

| | | | |
|---|---|---|---|
| 0 483 10 7 34F    | 0 396 13 7 1C4B   | 0 375 14 7 2173   | 36 465 15 7 416D  |
| 0 330 11 7 46F    | 0 318 13 7 1D0B   | 0 204 14 7 23A3   | 0 504 15 9 453F   |
| 0 324 11 7 64F    | 0 267 13 7 1C33   | 0 348 14 7 26C3   | 36 498 15 9 4ACF  |
| 0 258 11 7 58F    | 0 225 13 7 1B13   | 0 114 14 7 3643   | 0 402 15 9 598F   |
| 0 339 11 7 747    | 0 177 13 7 1D13   | 36 513 14 7 219D  | 36 363 15 9 4F0F  |
| 0 294 12 7 92F    | 0 231 13 7 15A3   | 0 273 14 7 2C2D   | 0 387 15 9 4A77   |
| 36 765 12 7 8CF   | 0 210 13 9 195F   | 0 87 14 9 259F    | 0 351 15 9 4D57   |
| 0 285 12 7 B17    | 0 228 13 9 1A5F   | 36 327 14 9 2977  | 0 291 15 9 7617   |
| 0 162 12 7 E2B    | 0 306 13 9 171F   | 36 369 14 9 23B7  | 0 159 15 9 47A7   |
| 0 357 12 7 B4B    | 0 249 13 9 166F   | 0 153 14 9 3CA7   | 36 537 15 9 5D27  |
| 0 171 12 7 A73    | 0 243 13 9 1DA7   | 0 303 14 9 2EC7   | 0 252 15 9 547B   |
| 0 360 12 7 D93    | 0 405 13 9 1F27   | 72 825 14 9 3E87  | 0 333 15 9 786B   |
| 0 423 12 7 CE3    | 0 180 13 9 173B   | 0 261 14 9 2C7B   | 0 417 15 11 5CBF  |
| 36 576 12 9 ADF   | 0 393 13 9 179B   | 36 357 14 9 3A9B  | 0 237 15 13 6EFF  |
| 36 642 12 9 CF7   | 0 147 13 9 1E9B   | 0 321 14 9 3A73   | 0 240 16 7 9847   |
| 0 234 13 7 10CF   | 36 492 13 11 1DDF | 0 276 14 11 35DF  | 0 279 16 7 816B   |
| 36 528 13 7 1587  | 0 141 13 11 1F6F  | 0 735 15 7 444F   | 0 195 16 7 912B   |
| 0 288 13 7 1C87   | 36 519 13 11 1FAF | 36 201 15 7 4457  | 36 783 16 7 8CA3  |
| 0 183 13 7 115B   | 0 213 14 7 2A27   | 0 96 15 7 4687    | 0 297 16 7 E843   |
| 36 630 13 7 146B  | 36 441 14 7 2A4B  | 36 438 15 7 4353  | 36 531 16 7 88CD  |
| 0 312 13 7 132B   | 0 432 14 7 2B0B   | 0 189 15 7 50E3   | |

*The string format: Gamma, Beta, constraint length, number of ones, polynomial in hexadecimal form.